\documentclass[epj]{svjour}

\usepackage{graphics}

\begin{document}

\title{Heterogeneity, quality, and reputation in an adaptive recommendation model}
\author{G. Cimini\inst{1} \and M. Medo\inst{1} \and T. Zhou\inst{1,2,3}\and D. Wei\inst{1} \and Y.-C. Zhang\inst{1}}
\institute{Physics Department, University of Fribourg, CH-1700 Fribourg, Switzerland \and 
Web Sciences Center, University of Electronic Science and Technology of China, Chengdu 610054, P. R. China \and 
Department of Modern Physics, University of Science and Technology of China, Hefei 230026, P. R. China}

\date{Received: date / Revised version: date}

\abstract{Recommender systems help people cope with the problem of information overload. 
A recently proposed adaptive news recommender model~[Medo et al., 2009] 
is based on epidemic-like spreading of news in a social network. 
By means of agent-based simulations we study a ``good get richer'' feature of the model and determine which attributes are necessary for a user to play a 
leading role in the network. We further investigate the filtering efficiency of the model as well as its robustness against 
malicious and spamming behaviour. We show that incorporating user reputation in the recommendation process 
can substantially improve the outcome.
}
\maketitle

\section{Introduction}\label{intro}
We live in an information-rich world with a vast number of sources competing for our attention~\cite{goldhaber,huberman1}.
In addition to the old-fashioned information distribution systems, such as newspapers, which favor news of very general
interest, recommender systems~\cite{resnick1,herlocker,adomavicius} act as personalized information filters 
by analyzing users' profiles and past activities. Techniques used to produce recommendations
include correlation-based collaborative filtering~\cite{herlocker,linden}, Bayesian clustering~\cite{breese}, 
probabilistic latent semantic analysis~\cite{hofmann}, matrix decomposition~\cite{maslov}, and many others. 
However recent works show that similarity of past activities often plays a less important role 
than social influence and recommendations obtained purely by abstract mathematical analysis are valued less 
than those coming from our friends or peers~\cite{sinha}. A new approach, \emph{social recommendation}, 
has hence emerged to make direct use of social connections between members of a society~\cite{goldbeck}.
Examples of popular implementations of social recommender systems include blogger.com and delicious.com, 
where each user can select some other users as information sources and imports blog articles or URLs 
from them. In these systems, information favored by an individual user spreads to the user's followers and, 
if favored again, to followers' followers, resembling an epidemics or rumor spreading in a network~\cite{zhou1,moreno}.

A recently proposed news recommender model mimics the spreading process typical for social systems 
and combines it with an adaptive network of connections~\cite{matus}. In this model, when a user reads a news, 
she can either ``approve'' or ``disapprove'' it. If approved, the news spreads to \emph{followers} of the given user 
(whom we refer to as \emph{leader}). Each user has an evolving set of leaders (or, according to the terminology 
of the original paper, sources) and can become a leader for other users. Simultaneously with spreading of news, 
the leader-follower network evolves with time to best capture similarity of users. In~\cite{matus} they provide a
detailed description of an agent-based approach which is used to assess model's behaviour and test its performance.

Every recommendation method, if it is to be implemented in real applications, has to respect the heterogeneity of users. 
Users may differ, for example, by how often they use the recommender system, how broad are their interests, 
and how accurate they are in evaluation of recommended news. In this work we study the effects of introducing user 
heterogeneity in the above-described adaptive recommendation model. 
We show that when frequency of being active and evaluation noise vary among users, leaders with exceptionally 
high numbers of followers appear and a scale-free-like leadership structure emerges. 
Scale-free networks are observed in diverse systems~\cite{caldarelli1} and over the past two decades 
they attracted considerable attention. The mechanism of their emergence based on user heterogeneity in a 
social recommendation process is similar to the previously discussed ``good get richer'' phenomenon~\cite{caldarelli2}.

Heterogeneity also means that some users may try to intentionally misguide the system by providing wrong evaluations. 
We study whether the system is robust against such malicious behaviour, and if it can suppress low-quality content 
and promote high quality. While the original adaptive recommender model already exhibits 
a notable resistance to malicious behaviour, we further improve it by introducing a simple measure of user reputation 
and employ a hybrid recommendation mechanism which combines similarities of users' rating patterns with reputation 
(for a review of reputation systems, see~\cite{resnick2}). 
We show that these changes enhance the filtering efficiency of the system and its robustness against various kinds 
of malicious behaviour, leaving its performance almost unchanged. The proposed combination of reputation 
and personalized recommendation hence seems as a promising candidate for real life applications.

\section{Description of the model}\label{sec.model}
We first briefly recall the original adaptive recommendation model introduced in~\cite{matus}. 
The system consists of $U$ users. Each user is connected to $L$ other users (to whom we refer as the user's leaders); 
in the network representation this corresponds to a monopartite directed network with $U$ nodes of fixed in-degree $L$. 
Evaluation of news (or a different kind of content) $\alpha$ by user $i$, $e_{i\alpha}$, is either $+1$ (liked), $-1$ (disliked) or $0$ (not rated yet).
Similarity of reading tastes of users $i$ and $j$, $s_{ij}$, is estimated by comparing past users' assessments. 
If $i$ and $j$ evaluated $N_{ij}$ common news and agreed in $A_{ij}$ cases,
their similarity can be measured in terms of the overall probability of agreement
\begin{equation}\label{eq.s}
s_{ij}=\frac{A_{ij}}{N_{ij}}\left(1-\frac1{\sqrt{N_{ij}}}\right)
\end{equation}
where the term in parentheses disadvantages user pairs with small overlap $N_{ij}$ (their similarity estimates, 
albeit possibly high, are prone to statistical fluctuations). If $N_{ij}=0$ then $s_{ij}$ is undefined and replaced by a small positive value $s_0$. 
Apart from their ratings, no other information about users is assumed by the model. 

Propagation of news is governed by their recommendation scores. We denote as $R_{i\alpha}$ the recommendation score 
of news $\alpha$ for user $i$. When news $\alpha$ is introduced to the system by user $i$ at time $t_{\alpha}$, its initial 
recommendation score is $R_{j\alpha}(t_{\alpha})=s_{ij}$ for users $j$ who are followers of $i$ and it is zero for the 
others (i.e., it cannot be recommended to them yet). 
In this way, the news is passed from user $i$ to $i$'s followers. If this news is later liked by one of users $j$ 
who received it, it is similarly passed further to this user's followers, and so on. A user may receive the same news 
from multiple leaders---recommendation scores are summed up in that case, reflecting that a news liked by several leaders 
is more likely to be liked by this user too. To allow fresh news to be accessed fast, recommendation scores are 
exponentially damped with time. In this way, novelty of news fades with an exponential law~\cite{huberman2}. 
By combining the described processes, we have the formula for the recommendation score
\begin{equation}\label{eq.R}
R_{j\alpha}(t)=(1-\delta_{|e_{j\alpha}|,1})\lambda^{t-t_{\alpha}}\sum_{l\in L_j}s_{jl}\:\delta_{e_{l\alpha},1}.
\end{equation}
Here $L_j$ is the set of leaders of user $j$ and $\lambda\in(0,1]$ is the damping factor. 
The term $\delta_{e_{l\alpha},1}$ is one when user $l$ liked news $\alpha$ and zero otherwise.
Similarly, the term $1-\delta_{|e_{j\alpha}|,1}$ equals one only when user $j$ has not rated news $\alpha$ yet. 
For user $j$ at time $t$, news are recommended according to their current score $R_{j\alpha}(t)$ (the higher, the better).
Note that the described damping mechanism is different from the one proposed in~\cite{matus} 
where the damping factor was additive and the damping occurred only if too many news were recommended to a user. 
Our motivation for decreasing scores always is that news lose their novelty regardless of being recommended or not, 
and that the multiplicative factor keeps $R_{i\alpha}(t)$ positive, hence even old news can be in principle read by users 
if there are no relevant fresh news with higher recommendation scores. Since the spreading of a news over a long path may take long time, 
recommendation scores decreasing with time not only enhance novelty in the system but also promote news that come from the local neighborhood,
effectively working as a local news filter.

Starting from an initial random network configuration (random assignment of leaders to users), connections are 
periodically rewired to drive the system to an optimal state where users with high similarity (taste mates) are directly 
connected. In this way the topological evolution of the network and the dynamics of the network's nodes becomes invariably linked, 
as in other adaptive co-evolutionary networks~\cite{gross}. Thus the updating procedure is an important part of the model. 
Some simple methods are:
\begin{enumerate}
\item\emph{Global rewiring.} Leaders are selected using all currently available information: for each user $i$, 
$L$ leaders with the highest similarity values $s_{ij}$ are selected. This is the best performing method but it is also 
computationally expensive as it requires computation of all $U(U-1)/2$ similarity values.
\item\emph{Random rewiring.} For each user, the leader with the lowest similarity value is replaced with a randomly chosen 
user (if this user is even less similar, no replacement occurs). This is the simplest possible method but its rate of 
convergence to the optimal state is, as we will see, very slow.
\item\emph{Local rewiring.} For each user $i$, the leader with the lowest similarity value is replaced with the most similar 
user among leaders of $i$'s leaders (hence we are exploring $i$'s neighborhood within the distance of two). This mechanism 
is based on the simple observation that two users who share a common neighbor are likely to be similar 
(for more sophisticated methods for link prediction in networks based on propagation of trust/similarity, see~\cite{guha,leskovec}). 
Computational cost of this method scales as $O(UL^2)$ and hence as long as $L^2<U$ (a mild constraint, since $L$ is small), 
this method is faster than global rewiring.
\item\emph{Hybrid rewiring.} Random rewiring is used in $10\%$ of cases and local rewiring is used in the others. 
This rewiring mimics the natural evolution of communities where users search for friends among friends of friends 
(local rewiring) but also casual encounters occur and may lead to long-term relationships (random rewiring).
\end{enumerate}
While the first three methods were already studied in \cite{matus}, the last one is novel.

For numerical tests of the model, we use the agent-based framework described in~\cite{matus}. 
Taste of user $i$ is represented by a $D$-dimensional binary vector $\mathbf{t}_i$ and attributes of news $\alpha$ by a 
$D$-dimensional binary vector $\mathbf{a}_\alpha$. Each vector has a fixed number, $D_A$, of elements equal one 
(active tastes) and all remaining elements equal zero. We always set the system so that all mutually different user 
taste vectors are present exactly once, hence $U={D \choose D_A}$. This also means that taste vectors of two users differ 
at least in two elements. Opinion of user $i$ about news $\alpha$ is based on the overlap of the user's taste vector 
with the news's attribute vector
\begin{equation}
\label{eq.O}
\Omega_{i\alpha}=(\mathbf{t}_i,\mathbf{a}_\alpha)
\end{equation}
where $(\cdot,\cdot)$ is a scalar product of two vectors. 
If $\Omega_{i\alpha}\geq\Delta_i$ user $i$ likes news $\alpha$ ($e_{i\alpha}=+1$), otherwise she dislikes it ($e_{i\alpha}=-1$). 
The value $\Delta_i$ is an approval threshold of user $i$; the higher it is, the more demanding the user is.

Simulation runs in discrete time steps. In each step, an individual user is active with probability $p_A$. 
When active, the user reads and evaluates the top $R$ news from her recommendation list and with probability $p_S$ 
submits a new news with attributes identical to the user's tastes. To save computational time, the network of connections 
is rewired every ten time steps. Finally to measure the system's performance, we use \emph{approval fraction} which is 
the ratio of approvals to all assessments and tells us how often users are satisfied with the news they get recommended, 
and \emph{average differences} which is the average number of vector elements in which users differ from their leaders 
and tells us how well the network has adapted to users' tastes. 

\subsection{Rewiring performance}
\begin{figure}
\resizebox{0.47\textwidth}{!}{\includegraphics{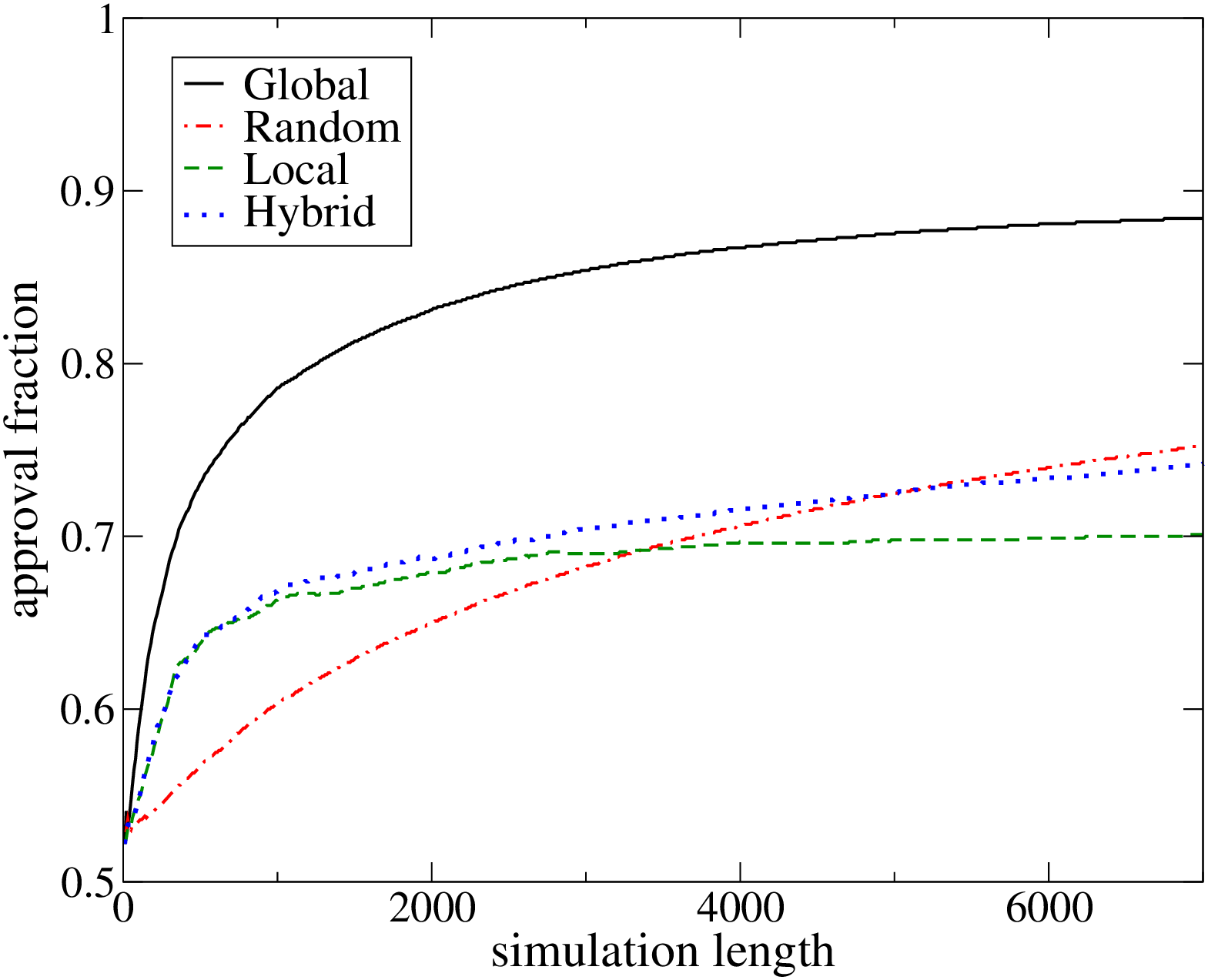}}\\[4pt]
\resizebox{0.47\textwidth}{!}{\includegraphics{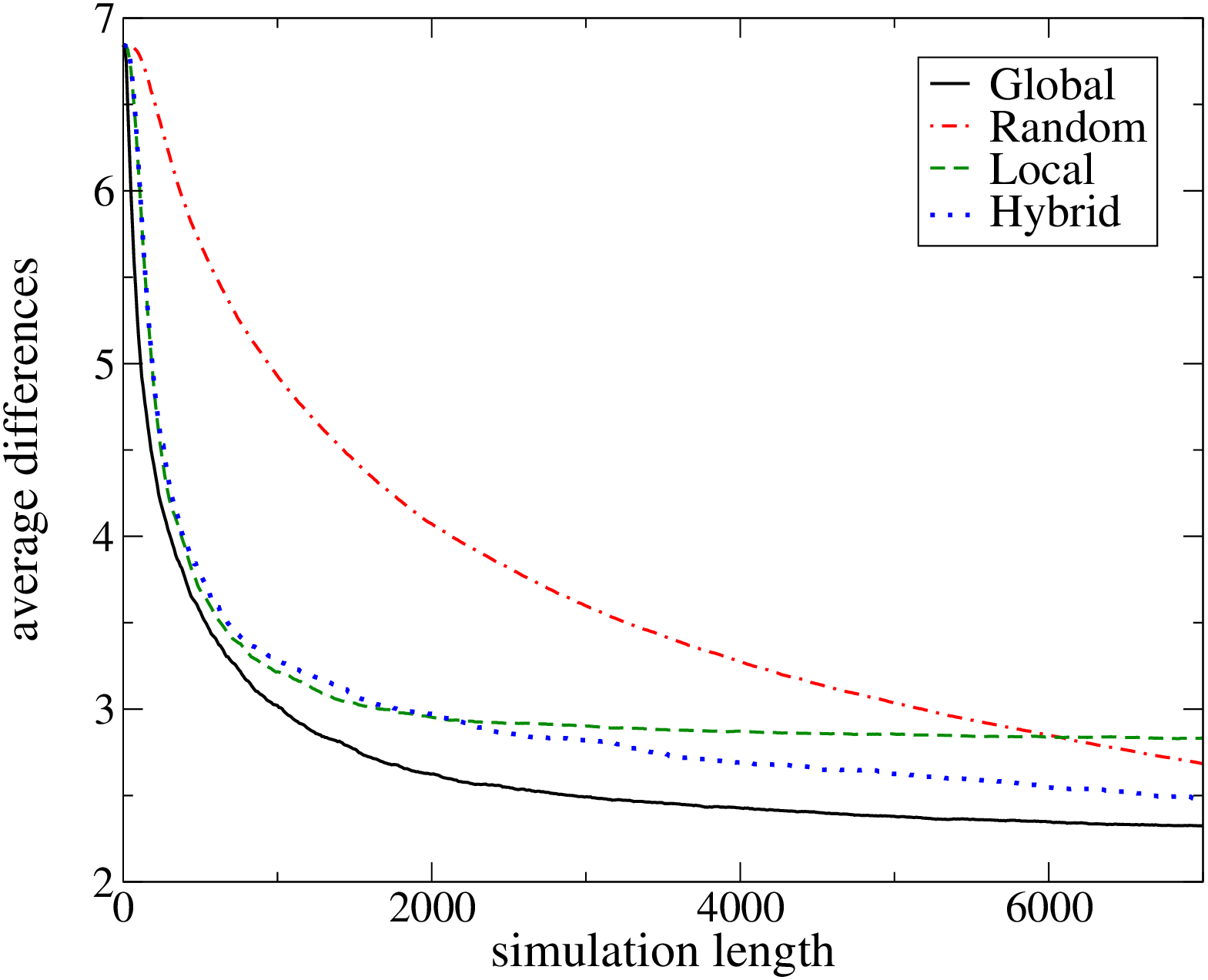}}
\caption{Comparison of rewiring mechanisms for $D=14$, $D_A=6$, $L=10$, $R=3$, $p_A=0.05$, $p_S=0.02$, $\Delta=3$, $\lambda=0.9$.}
\label{fig.AFD}
\end{figure}

Since the aforementioned hybrid rewiring method is new, we begin this study with its comparison to the previously known 
methods. For simplicity we assume a homogeneous setting of users with identical values of $\Delta_i$, $p_A$, and $p_S$. 
Figure \ref{fig.AFD} shows that all methods are able to gradually improve both approval fraction and average differences. 
Apart from local rewiring, the other three methods slowly approach the optimal assignment of leaders with average 
differences equal two. This ability to converge is due to a gradually increasing pool of commonly evaluated news 
which allows for precise similarity estimates and, eventually, the optimal assignment of leaders. Since $p_A$ is small, the amount of 
available information grows slowly and employing the rewiring more often would not make the convergence much faster. 
Note that as for each user there are $N=D_A(D-D_A)$ possible optimal leaders who differ exactly in two taste elements, the 
optimal state is unique only if $L=N$. 
If $L<N$ there are different possible optimal states which are equivalent in term of global properties of the system. 
Initial conditions and users' dynamics determine the particular equilibrium state of the system. 
If $L>N$ (which is not our case, however), average differences are greater than two even in the optimal state. 

By contrast, local rewiring reaches only a sub-optimal assignment of leaders (the degree of sub-optimality strongly
depends on the ratio between the number of optimal leaders to the total number of users, and also on the particular realization of system's evolution). 
This is because if the network's 
evolution once stops in a sub-optimal state, there is no means to escape from it with local rewiring: if user's best 
taste mates are at that moment out of the second-order neighborhood, they can never be reached. 
In other words, the effectiveness of local rewiring is limited by the current network's topology, which completely determines 
the pool of candidate leaders for each user. Unlike other rewiring methods, such pool is very small compared to the whole network 
($\leq L^2$ users) and it changes slowly in time. This trapping in a sub-optimal state is hence similar to the trapping of greedy 
optimization algorithms in a local minimum. 

\begin{figure}
\resizebox{0.47\textwidth}{!}{\includegraphics{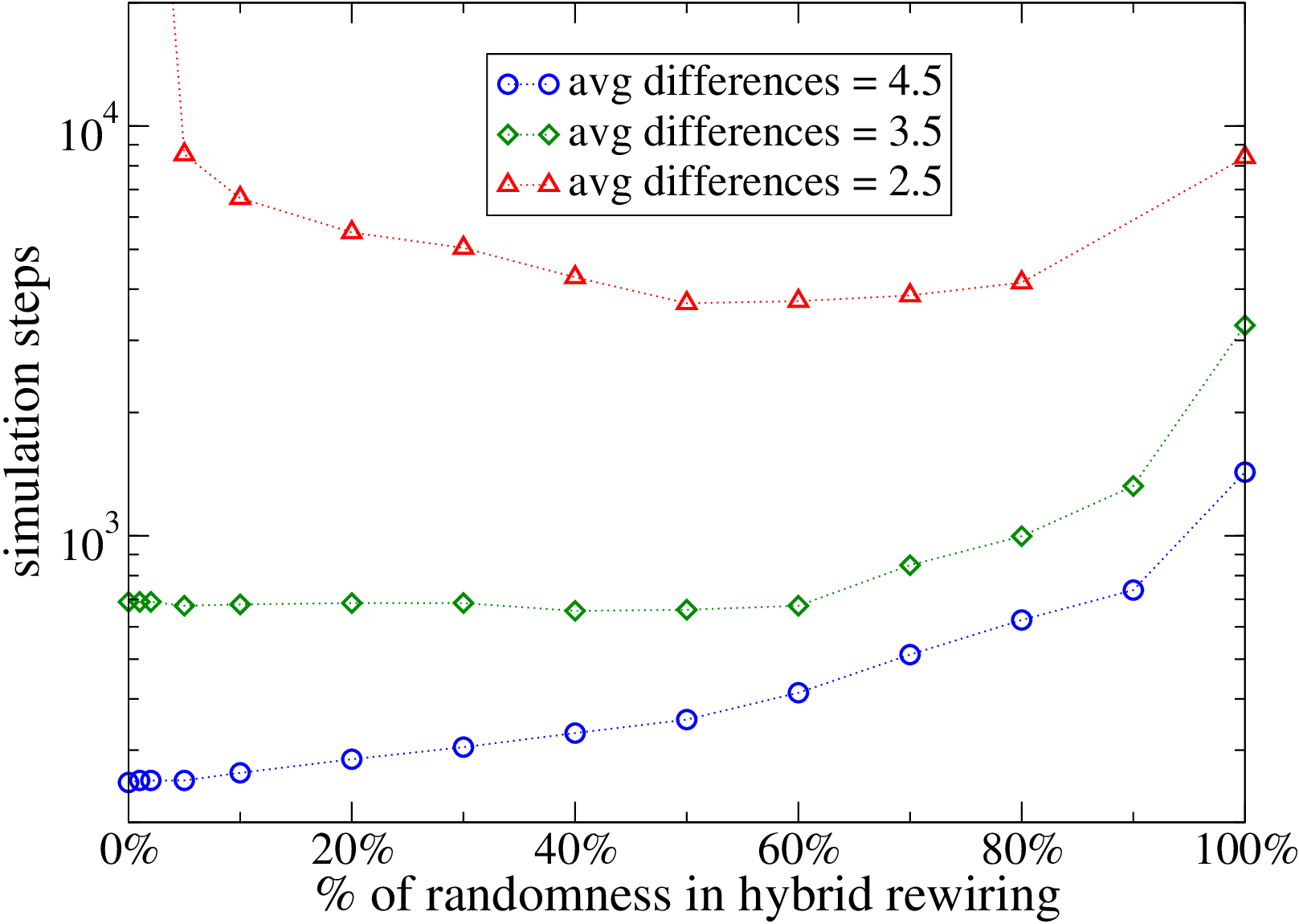}}
\caption{
Simulation steps needed by hybrid rewiring---with different percentage of randomness---to reach different values of the average differences in the system. 
Simulation parameters as in Figure \ref{fig.AFD}.
At the beginning of the evolution the local facet of the rewiring speeds up the convergence (see the line relative to average differences equal to 4.5). 
However, as system's approach equilibrium, the random facet becomes more suitable for network's exploration, 
as it allows to connect users regardless of their distance (see line relative to average differences equal to 2.5). 
We employ a randomness of 10\% to have both fast convergence at the beginning of the evolution and reasonable time to get to the ground state.}
\label{fig.HrP}
\end{figure}

Methods' convergence rates differ significantly, with global and random rewiring being the fastest and slowest, respectively. 
Notably, the hybrid method converges almost as fast as the global one (the relation between system's convergence rate and the percentage of randomness 
used in the hybrid rewiring is shown in Figure \ref{fig.HrP}). We conclude that hybrid rewiring represents 
a favorable compromise~between performance
and computational complexity, hence it is used in all following simulations.

\section{Heterogeneity and leadership}\label{sec.hetero}
In real social networks there are people with different profiles. In this section we study the effects of usage frequencies 
and judgment abilities on the leader-follower network. Activity frequencies $p_A$ are drawn from a power-law distribution
\begin{equation}
\label{eq.act}
P(p_A)\sim p_A^{-\gamma},\quad p_A\in[\eta,1].
\end{equation}
In this way we obtain a very diverse set of activity frequencies which mimics the observed scale-free patterns 
of human behaviour~\cite{zhou2}. Exponent $\gamma$ can be tuned to obtain a desired percentage of highly active users. 
In our simulations we set $\eta=0.01$ and $\gamma=2$ which implies that $10\%$ of users have $p_A>10\eta$. 
For the sake of simplicity we assume $p_S=p_A/10$ (that is, a user who is often online also has a high submitting rate). 
This assumption gets on well with real life experience: high usage users are also the ones who contribute most 
to the functioning of the system by introducing hot news. The second source of user heterogeneity lies in diverse levels 
of errors present in their evaluations. We model this feature by generalizing equation~(\ref{eq.O}) to
\begin{equation}
\label{eq.Ox}
\Omega_{i\alpha}=(\mathbf{t}_i,\mathbf{a}_\alpha)+ux_i
\end{equation}
where $u$ is a random value drawn at each assessment from the uniform distribution with domain $[-1,1]$ 
and $x_i$ is the fixed magnitude of evaluation errors for user $i$, distributed uniformly in $[0,X]$.

\begin{figure}
\resizebox{0.47\textwidth}{!}{\includegraphics{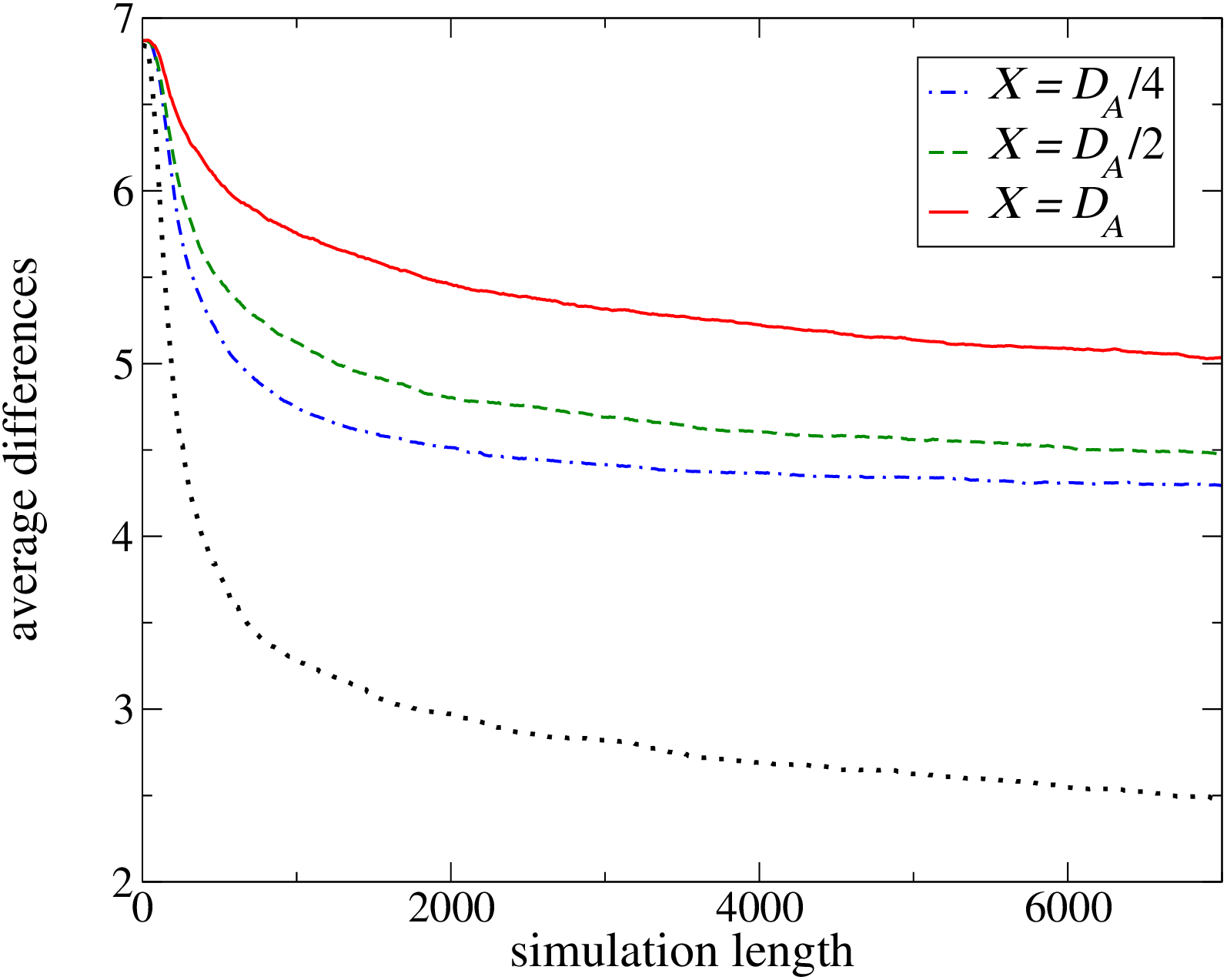}}\\[4pt]
\resizebox{0.47\textwidth}{!}{\includegraphics{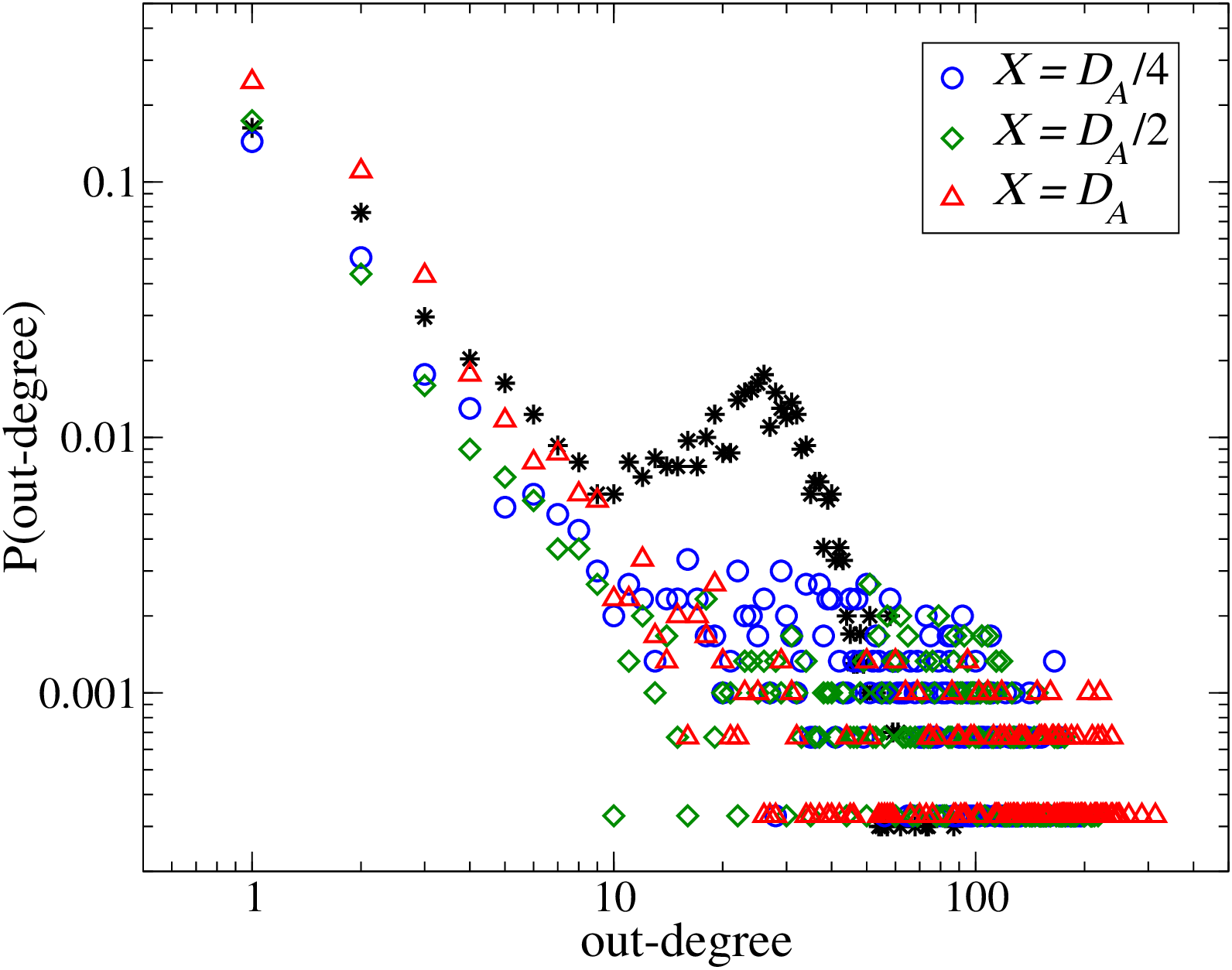}}
\caption{Average differences vs time (upper panel) and out-degree distribution (bottom panel) for heterogeneous settings 
with various values of maximal magnitude of evaluation noise $X$. For comparison, results for the original homogeneous 
setting (with $X=0$, $p_A$ equal to the average activity in the homogeneous setting and $p_S$ such to have 
the same average number of news in the system) is shown with the dotted line (upper panel) and stars (bottom panel). 
Parameter values as in Figure~\ref{fig.AFD}. Bottom panel also reveals the spontaneous emergence of two 
classes of users (with high and low out-degree respectively), a phenomenon typical of adaptive networks~\cite{ito}.}
\label{fig.plx_hyb}
\end{figure}

Figure~\ref{fig.plx_hyb} illustrates the impact of heterogeneity on the system. 
The upper panel shows the time evolution of the network. Compared to the original homogeneous case 
(which is shown with a dotted line), convergence to the optimal state is lost and the evolution itself 
is so slow that the system can be considered to stay in a quasi-steady and sub-optimal state.
Moreover, as shown in the bottom panel, the out-degree distribution (recall that a user's out-degree is equal to the number 
of the user's followers) becomes very broad. The initial part of the distribution can be fitted by a power law with 
exponent approximately $1.5$. A similar distribution arises also when global rewiring is used, 
though it is narrower than in the case of hybrid rewiring (the corresponding power-law exponent rises to approximately 
$2.0$). This suggests that the emergence of a scale-free leadership structure is related to self-organization 
in the society~\cite{cano} and that a centralized control favors more homogeneous resulting states. 

System dynamics can be explained by the presence of users who have high usage frequencies and, in turn, also much more 
evaluations of news than the average. At the beginning of the evolution, a large overlap of users' rating histories favors 
the formation of links (this feature does not depend on the term in parentheses in (\ref{eq.s})), and high usage users are 
obviously in advantage: they quickly attract many followers and become hubs of the network. Then if two taste-mates are 
linked to different hubs, even as time runs further they rarely evaluate the same news and their high similarity remains 
undiscovered: connections with high usage users are not abandoned and the network is trapped in a sub-optimal state and 
cannot evolve further. A high submitting rate for high usage users magnifies this phenomenon, as there are much more news 
which precisely reflect the tastes of these users (with a constant value of $p_S$ for each user, the convergence of the system is 
improved and the out-degree distribution gets slightly narrower). 
Evaluation noise plays an important role as well: precise users are preferred as leaders because they forward news 
that really match their tastes (and thus probably also followers' tastes). Besides, as these users give accurate ratings, 
they get a more stable and reliable similarity score with other users than the average, hence are easier to be identified 
as taste mates by other users. On the other hand, evaluations of a user with a very large error magnitude are basically 
random and hence the resulting similarity with any other user is close to $0.5$ which makes this user unlikely 
to be selected as a leader. 

\begin{figure}
\centering
\resizebox{0.42\textwidth}{!}{\includegraphics{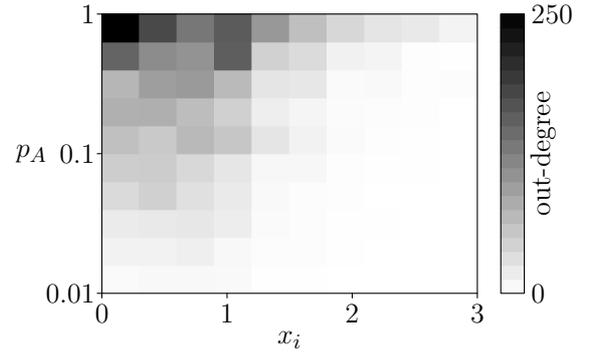}}
\caption{Out-degree versus activity frequency $p_A$ and individual evaluation error magnitude $x_i$ 
($X=D_A/2$, other parameters as in Figure~\ref{fig.AFD}).}
\label{fig.ODplx}
\end{figure}

Figure \ref{fig.ODplx} reports how both usage frequency and evaluation noise affect user's out-degree. 
As explained above, highly active users and precise users have on average more followers than other users. 
Note that active but imprecise users, as well as precise but lazy ones, cannot be popular leaders as opposed to the few 
who posses both features. These exceptional users attract a large number of followers, allowing for the scale-free 
leadership structure to emerge. This behavior is similar to the ``good get richer'' mechanism~\cite{caldarelli2} which explains 
a scale-free network structure on the basis of intrinsic fitness values of nodes. 

\section{Quality and reputation}\label{sec.QR}
\begin{figure}
\resizebox{0.47\textwidth}{!}{\includegraphics{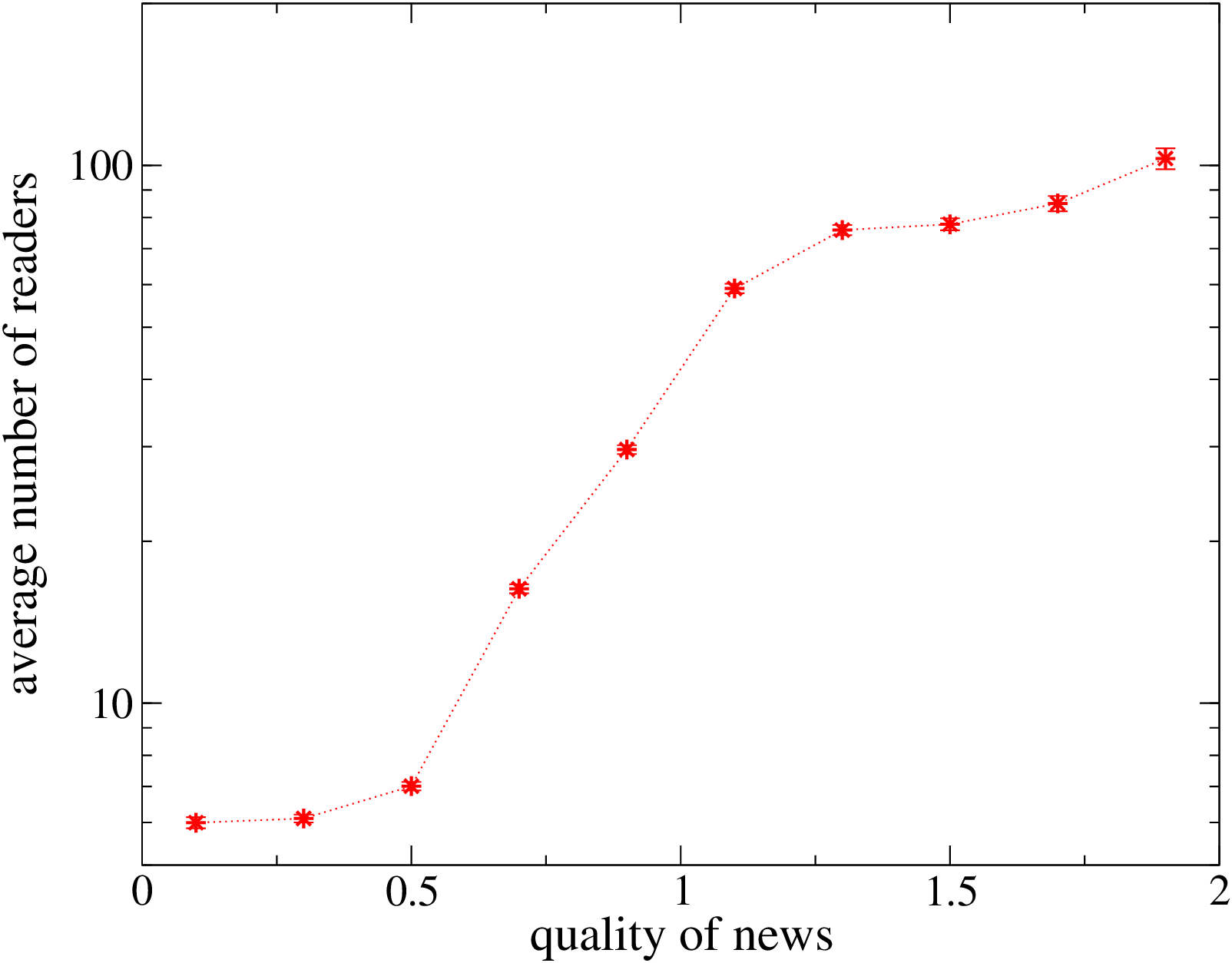}}
\caption{The dependency between the number of readers and the news' quality. Simulation parameters: 
$D=14$, $D_A=4$, $L=5$, $R=10$, $p_A=0.05$, $p_S=0.1$, $\Delta=2$, $\lambda=0.9$, $X=D_A/4$, 
hybrid rewiring of the network every 50 time steps.}
\label{fig.readq}
\end{figure}

Until now it was only the overlap between user's tastes and news's attributes what distinguished a liked news from 
a disliked one. Now we shall amend the rating process by another important factor, intrinsic quality of news. 
To this end, we assign a real-valued quality $Q_{\alpha}$ to each submitted news and generalize 
equation (\ref{eq.O}) to the form
\begin{equation}
\label{eq.OQ}
\Omega_{i\alpha}=Q_\alpha\cdot(\mathbf{t}_i,\mathbf{a}_\alpha).
\end{equation}
Quality of news is chosen when the news enters the system and does not change with time.\footnote{The quality factor in (\ref{eq.OQ}) 
transforms the overlap from integer to real value, resulting in a smoother 
dependence of system behaviour on approval threshold $\Delta$. Introduction of $Q_{\alpha}$ hence makes simulation results 
more robust and easier to analyze.} We draw $Q_{\alpha}$ from the normal distribution with mean $1$ and 
standard deviation $1/2$ (normal distribution is chosen to have only a small number of exceptionally good or bad news); 
when $Q_{\alpha}$ lies out of the range $[0;2]$, the draw is repeated. Figure~\ref{fig.readq} shows how news 
of different qualities propagate over the network. Remarkably, the recommender system has a high filtering efficiency: 
high-quality news spread to many users while low-quality news perish quickly. 
Saturation of the number of readers for high-quality news is mainly due to the damping factor $\lambda$. 
We remark that the spreading of a news in the system can be compared to a branching process~\cite{lorenz} of the number of the news' readers. 
News' propagation stops only when there are no users who could read or like it. Such a cascade can either die out quickly 
(when the news is liked by few) or invade a finite fraction of the system (when it is liked by many). 

Once we have introduced the concept of news quality to our simulations, it is straightforward to use it to investigate
system vulnerability to malicious behaviour. We introduce two different kinds of malicious users to our system: 
(a) users with \emph{non-informative} ratings (either rating at random, always liking, or always disliking), 
(b) \emph{spammers} who intentionally introduce low-quality content. Non-informative users are easily taken care of 
by the system because their similarity values with normal users are small and they are soon disconnected from the network. 
In particular, all-like and all-dislike users have high mutual similarity and hence they form small separate communities.
Our adaptive system is thus robust against malicious users of this kind.

With respect to spammers, the system is rather robust to their actions because a single low-quality news introduced 
by a spammer spreads only to a limited number of spammer's followers and as soon as they dislike the news, 
the news is removed from the system without affecting a large number of users. 
Alas, spammers can submit a large amount of worthless content and hence even a limited impact of each individual 
low-quality news can contribute to substantial discomfort of users. 
One could further argue that when followers of spammers dislike their low-quality news, spammers' similarity values
suffer and soon they are left with no followers. However, as we shall soon see, spammers can easily mask themselves 
by reasonably rating other news and hence keep their followers. At the same time, users submitting high-quality content 
are not rewarded with high popularity in the original model.

Instead of studying spammers and providers of good content, we pose a more general question: if the quality 
of submitted news differs from one user to another, what is the relation between the quality of news posted by a user 
and this user's out-degree? To simulate users with different submitting abilities we simply assume that each user 
has assigned a quality $Q_i$ and a news takes its quality from the user who submits it. In this way we obtain a system 
where some users always introduce low-quality content (spammers) and others who submit high-quality news (good sources).

We introduce \emph{reputation} as a tool to discriminate users. Reputation systems, already widely used 
in successful commercial online applications, represent an important class of decision support tools that can help reduce 
risk when engaging in interactions on the Internet and also encourage good behaviour~\cite{resnick2}. 
Reputation itself is a measure of trustworthiness based on referrals or ratings from other members of 
a community~\cite{josang,freeman}. In our case, we introduce the reputation score of user $i$ as
\begin{equation}
\label{eq.rep}
r_i=\frac{\sum_{\alpha\in I_i}l_\alpha}{|I_i|}
\left(1-\frac{1}{\sqrt{|I_i|}}\right)
\end{equation}
where $I_i$ is the set of news introduced by $i$ and $l_{\alpha}$ is the fraction of all users\footnote{It is also possible 
to define $l_{\alpha}$ using only the users who rated news $\alpha$. Numerical simulations show that using 
the former definition better distinguishes users with different $Q_i$.} who liked news $\alpha$; 
when $I_i=0$ (no news submitted by this user), we set $r_i=0$. Using user similarity and reputation, we set the strength 
of the link coming from user $j$ to user $i$ as
\begin{equation}
\label{eq.sr}
s_{ij}'=ms_{ij}+(1-m)r_j
\end{equation}
where $m$ is a mixing parameter which sets the weight of similarity and reputation in the recommendation process 
(notice that $s_{ij}'$ is not symmetric). This mechanism differs from the traditional \emph{popularity-based} 
recommendation in replacing the object's popularity with that of the author as well as in using a spreading mechanism 
in a social adaptive network. When $m=1$, we recover the original reputation-free model, when $m=0$,
recommendation is based purely on reputation, and submitters of news of general interest are favoured 
by achieving a high value of $l_\alpha$.
\begin{figure}
\resizebox{0.47\textwidth}{!}{\includegraphics{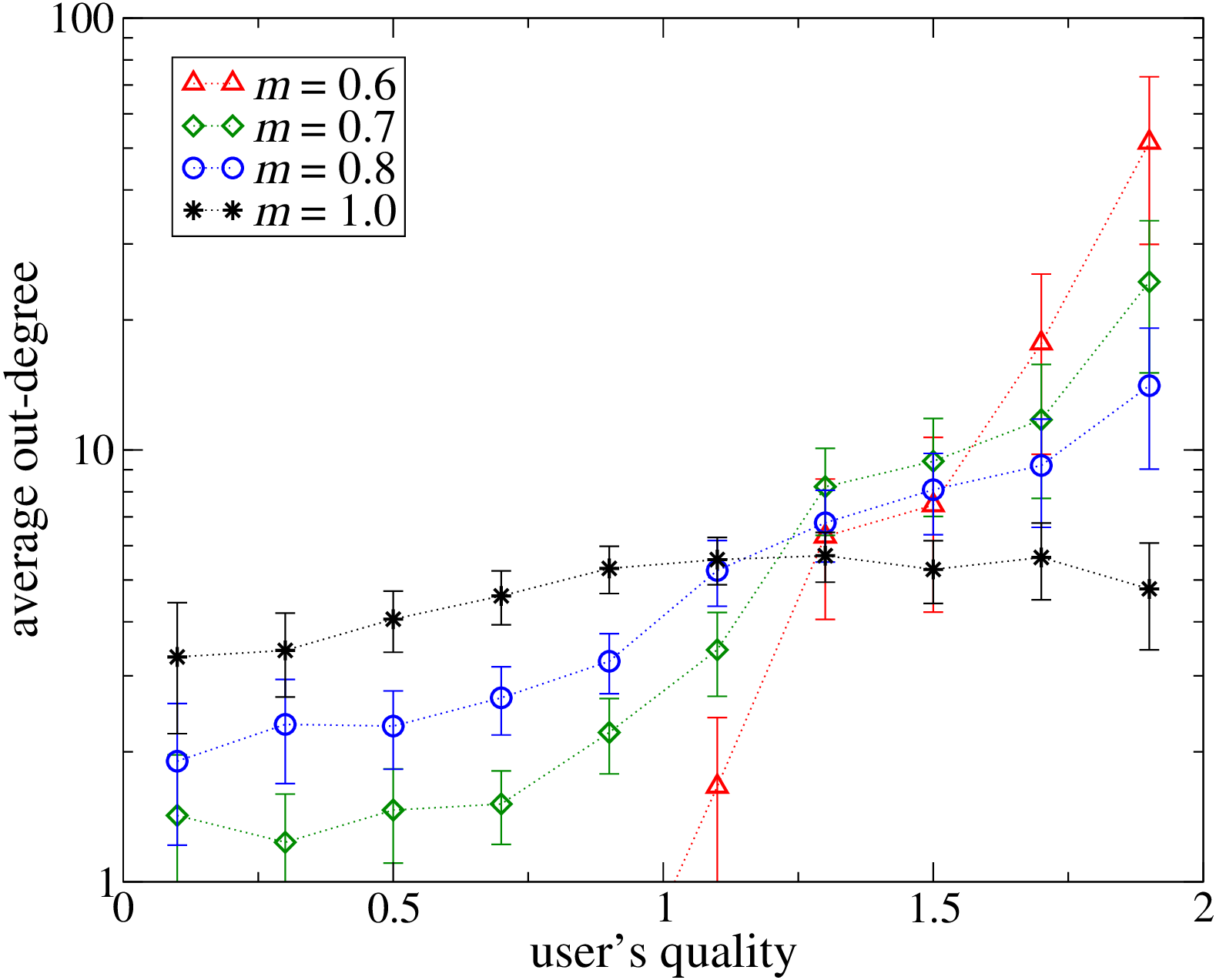}}\\[4pt]
\resizebox{0.47\textwidth}{!}{\includegraphics{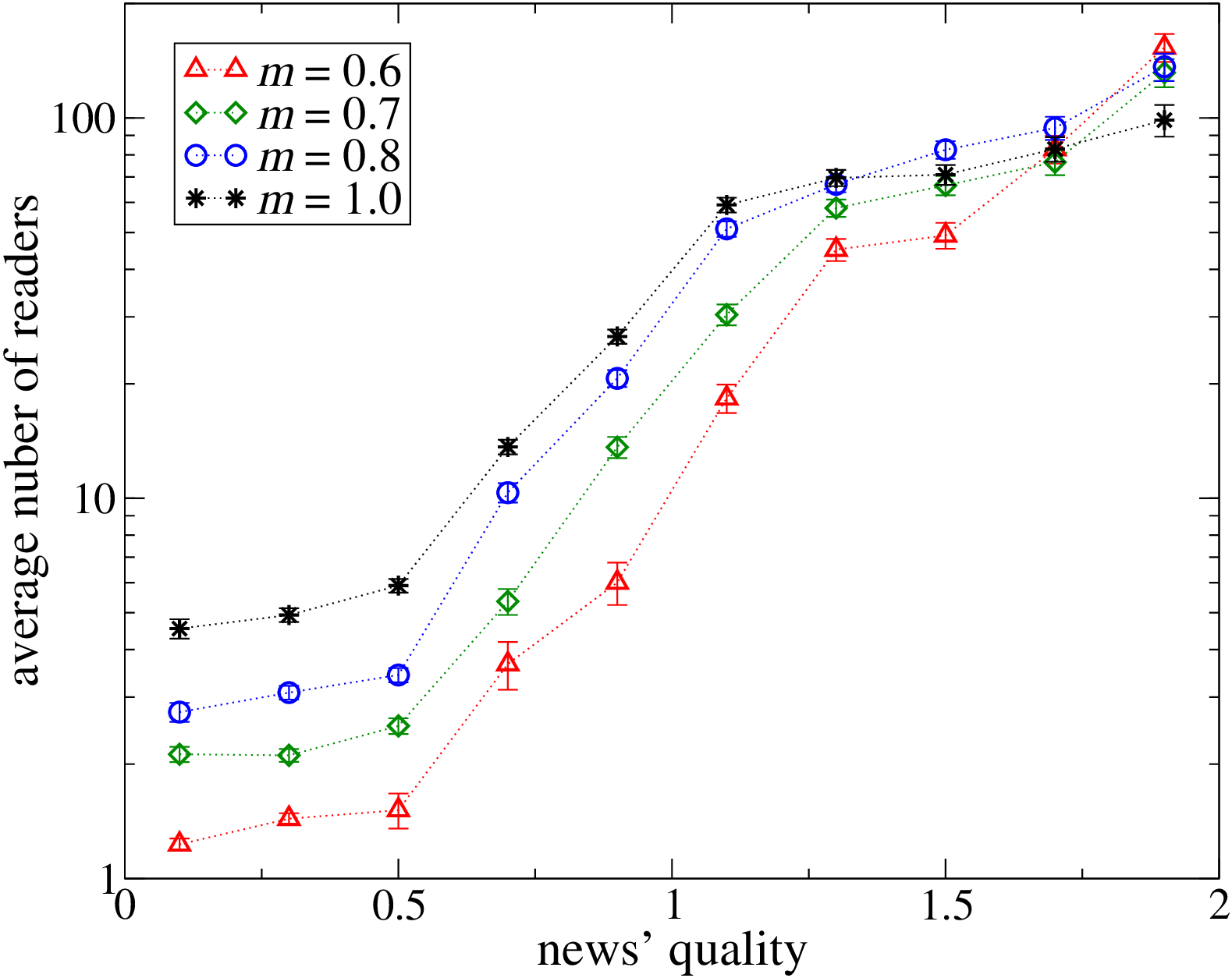}}
\caption{Number of followers vs user's quality (upper panel) and number of readers vs news's quality (bottom panel) 
for different values of $m$. Parameters values as in Figure \ref{fig.readq}.}
\label{fig.odeg_read_q}
\end{figure}
\begin{figure}
\resizebox{0.47\textwidth}{!}{\includegraphics{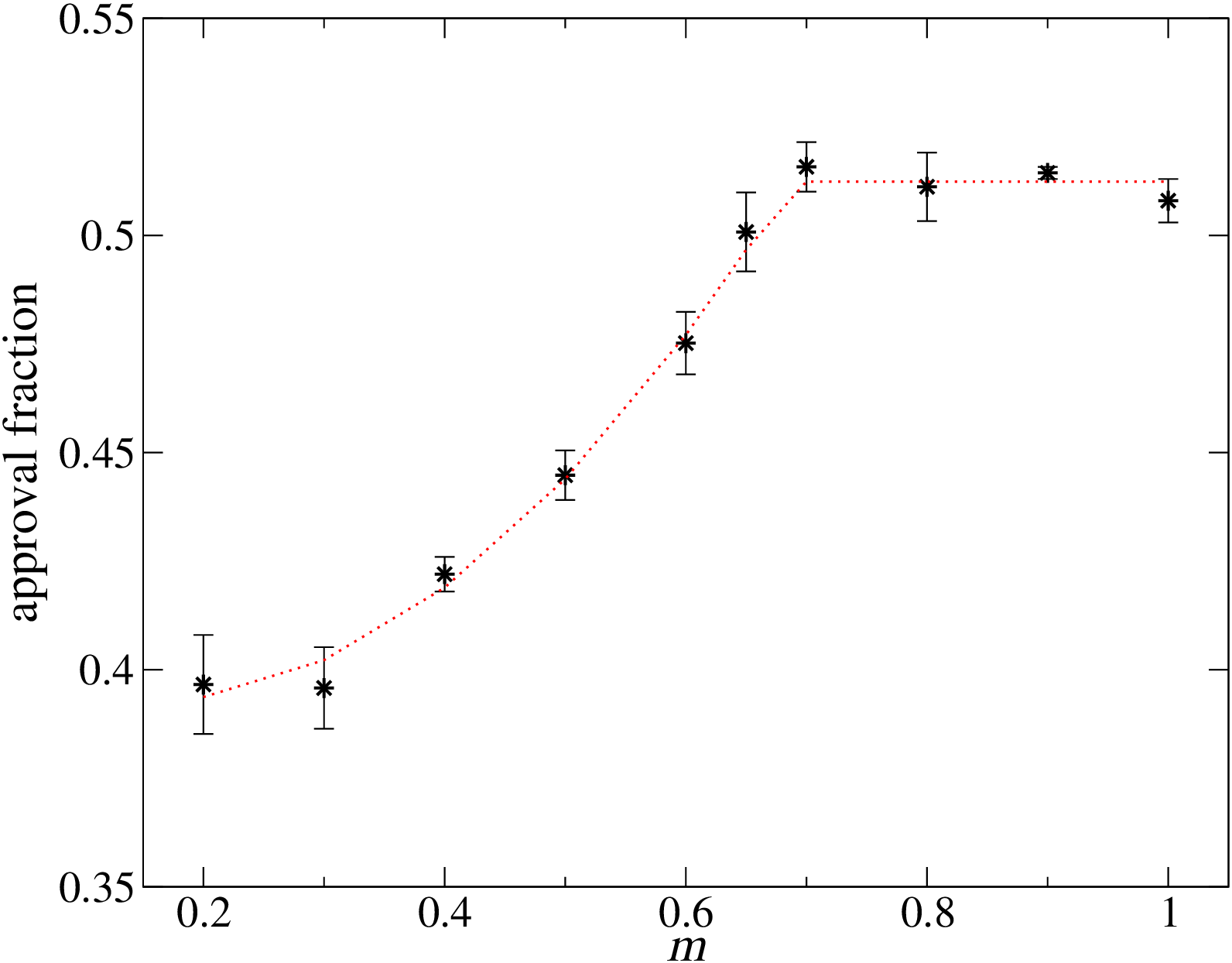}}
\caption{Stationary values of the approval fraction for different values of $m$. Simulation parameters as in Figure \ref{fig.readq}.}
\label{fig.af_q}
\end{figure}

In simulations we draw users' quality values $Q_i$ from the same distribution that we used for news' quality values
$Q_{\alpha}$ before. As shown in Figure~\ref{fig.odeg_read_q}, in the original setting ($m=1$) 
user's number of followers does not depend on the user's quality---a feature that has been discussed above. 
When $m$ is significantly less than $1$, reputation of users plays an important role and users with low values of $Q_i$ 
can be left with no followers (when $m\leq 0.7$). Moreover, as the introduction of the reputation system causes news' qualities to affect the recommendation scores, 
the relative size of cascades in news propagation is magnified. Therefore, the similarity-reputation hybrid mechanism increases 
the filtering capability of the system. When $m=0$, leaders are selected and news are recommended purely according 
to reputation. As a result, recommended news are diverse and of high quality but not personalized for each individual user. 
Thus when the role of reputation is too big ($m$ is too small) users' satisfaction decreases. 
This is reported in Figure \ref{fig.af_q} where, when $m$ is small, approval fraction is lower than in the original model. 
At $m\approx0.7$ we observe a behaviour which is similar to a second order phase transition: 
approval fraction suddenly stops to grow and remains practically constant until $m=1$. 
This stationarity of approval fraction, while somewhat surprising, in fact makes our system easier to tune: 
all values of $m$ between $0.7$ and $1.0$ are equally good (with respect to approval fraction) and hence we can freely
decide how much we want to suppress users providing low-quality content (\emph{cf.} Figure~\ref{fig.odeg_read_q}).

\section{Conclusion}\label{concl}
After the advent of Web 2.0, many on-line resource-sharing websites have been developed and their popularity 
grows steadily. Some of them (\emph{delicious.com}, \emph{douban.com}, and others) recently introduced 
social recommendation where users can recommend content to others and in turn receive recommendations for themselves. 
Fast growth of online communities~\cite{kumar} and users' preference for recommendations from friends~\cite{sinha}
make social recommendation a promising way to better organize and deliver online resources and to enhance 
users' experience as well as social contacts.

The news recommender model introduced in \cite{matus} and further analyzed and improved in this work 
mimics spreading processes in adaptive social networks. 
It makes use both of users submitting new content as well as of other users rating that content and deciding 
its future fate in the system. We studied the behaviour and performance of this model in artificial computer simulations. 
We proposed a new method for the network's adaptation. This method is almost as efficient as global optimization 
using all available information, yet it is computationally much less expensive. Investigation of user 
heterogeneity showed that users' personalities strongly influence the properties of the resulting leader-follower network 
and give rise to a ``good get richer'' mechanism which was suggested in previous theoretical studies of complex
networks~\cite{caldarelli2}. Our simulations show that popularity of individual leaders is very broadly distributed; 
it can be partially described by a power law with exponent around 1.5. We further studied model's resistivity 
against reckless and malicious behaviour of users. Although the original model is already rather resistant to such users, 
we showed that when user reputation is introduced and recommendations are obtained by mixing this reputation 
with user similarity, power of malicious users can be further lowered and diffusion of good contents in the system enhanced.

Agent-based models similar to the one studied here can contribute greatly to our understanding of 
social systems~\cite{miller} as they allow us to study the effect of each individual model's assumption 
on the simulation outcome. The drawback is that the complexity of assumptions can be such that it is hard to 
make a link between the model and the modeled system. In addition to our efforts to make results robust with respect 
to the assumptions, it still would be beneficial to have direct empirical input for user behaviour. We envision a real 
implementation of the studied recommendation model as an ideal source of this kind of information, 
serving as a useful tool for users and a unique living laboratory for researchers.

\begin{acknowledgement}
We acknowledge stimulating discussions with C.-H. Yeung. 
This work was partially supported by the Future and Emerging Technologies programmes of the European Commission 
FP7-ICT-2007 (project LiquidPublication, grant no. 213360) and FP7-COSI-ICT (project QLectives, grant no. 231200).
\end{acknowledgement}

\end{document}